\documentclass[prb,twocolumn,superscriptaddress]{revtex4}
\usepackage{graphicx}
\usepackage{dcolumn}
\usepackage{bm}

\begin{document}

\title{Strong rejuvenation in a chiral-glass superconductor}

\author{A. Gardchareon}
\affiliation{Department of Materials Science, Uppsala University, Box 534, SE-751 21 Uppsala, Sweden}
\affiliation{Faculty of Sciences, Chiang Mai University, Chiang Mai 50200, Thailand}

\author{R. Mathieu\cite{newaddress}}
\affiliation{Department of Materials Science, Uppsala University, Box 534, SE-751 21 Uppsala, Sweden}

\author{P. E. J\"onsson}
\affiliation{Department of Materials Science, Uppsala University, Box 534, SE-751 21 Uppsala, Sweden}

\author{P. Nordblad}
\affiliation{Department of Materials Science, Uppsala University, Box 534, SE-751 21 Uppsala, Sweden}

\date{\today}

\begin{abstract}
The glassy paramagnetic Meissner phase of a Bi$_2$Sr$_2$CaCu$_2$O$_x$ superconductor ($x$ = 8.18) is investigated by squid magnetometry, using ``dc-memory'' experiments employed earlier to study spin glasses. The temperature dependence of the zero-field-cooled and thermo-remanent magnetization is recorded on re-heating after specific cooling protocols, in which single or multiple halts are performed at constant temperatures. The 'spin' states equilibrated during the halts are retrieved on re-heating. The observed memory and rejuvenation effects are similar to those observed in Heisenberg-like spin glasses.
\end{abstract}

\maketitle

While superconductors exhibit a negative susceptibility, a positive field cooled magnetization has been observed in several high temperature superconducting systems when cooled in sufficiently low magnetic fields.\cite{pme,pme2,pme3}
This phenomenon is referred to as the paramagnetic Meissner effect (PME), and can in certain cases be related to the appearance of spontaneous magnetic moments below $T_c$, created by spontaneous orbital currents associated with superconducting loops containing an odd number of $\pi$-junctions\cite{pijunc,pme3} (i.e. Josephson junctions inducing a phase shift of $\pi$).

The PME has been observed in Bi-2212 materials prepared by various methods. An especially striking PME is found in melt-cast crystals\cite{pme2} of Bi$_2$Sr$_2$CaCu$_2$O$_x$ ($x$ = 8.18); these melt-cast samples\cite{tem} have densely packed large grains with a strongly polydomain microstructure which is assumed to favor the creation of $\pi$-loops. The low temperature paramagne\-tic Meissner (PM) state of such a material has been found to exhibit glassy dynamics, including aging and memory phenomena.\cite{evieprl,evieejb} In ordinary spin glasses, such as the Ising-like Fe$_{0.5}$Mn$_{0.5}$TiO$_3$ [\onlinecite{ito}] and the isotropic (Heisenberg-like) dilute magnetic alloys Cu(Mn) and Ag(Mn),\cite{Young,granberg} a random distribution of the magnetic interaction yields magnetic disorder and frustration. Similarly, the glassiness of the PM state of the melt-cast sample may originate from a network with a random distribution of 0- and $\pi$-junctions within and in-between the superconducting grains.\cite{KawJPS,kawamura2}

In the present paper, the glassy PM state is investigated by recording the zero-field cooled and thermo-remanent magnetization after specific cooling protocols. Earlier reported aging and memory characteristics\cite{evieprl,evieejb} are re-confirmed. The main focus of this investigation is put on the observation of strong rejuvenation effects quite similar to those observed in an isotropic Cu(Mn) or Ag(Mn) spin glass, in which the existence
of a spin glass phase is suggested to arise from a chiral glass transition. \cite{Kaw,Dor}

Magnetization measurements were performed on a granular Bi$_2$Sr$_2$CaCu$_2$O$_x$ sample, with $x$ = 8.18 (Bi2212), manufactured by a melt-cast process.\cite{pme2,tem}
The onset of superconductivity occurs at $T_c$ $\sim$ 87 K. The temperature dependence of the zero-field cooled (ZFC), field cooled (FC) and thermo remanent (TRM) magnetization was recorded on re-heating from the lowest temperature using a non-commercial low-field Superconducting QUantum Interference Device (SQUID) magnetometer.\cite{magnusson}
 The dc magnetic field (from 0.01 Oe to 1 Oe) employed in the experiments is generated by a small superconductive solenoid coil always working in persistent mode during measurements; a small background field of 0.6 mOe was compensated.  The magnetization is recorded while keeping the sample stationary, in the center of the third coil of a third order gradiometer.

In the case of the ZFC (resp. TRM) magnetization, temperature stops were performed during cooling. In these experiments, the sample was rapidly cooled ($\sim$ 4K/min) in zero magnetic field (resp. a small magnetic field) from a reference temperature $T_{\rm{ref}}$ = 95 K above $T_c$ down to a temperature $T_s$ below the critical temperature, at which a halt was made for some duration (1 000 s or 10 000 s). The cooling was then resumed down to the lowest temperature (60 K), and a small magnetic field was applied (resp. cut to zero). The magnetization was then recorded on re-heating (heating rate $\sim 1$~K/min). Experiments including two or more stops were performed. ZFC and TRM relaxation were also recorded at the stop temperatures for comparison. In the relaxation experiments, the sample was rapidly cooled from $T_{\rm ref}$ to the measurement temperature $T_m(= T_s)$ in zero magnetic field (ZFC case), and after a wait time, $t_w$, a small magnetic field was switched on, and the magnetization recorded with time.

\begin{figure}[htb]
\includegraphics[width=0.45\textwidth]{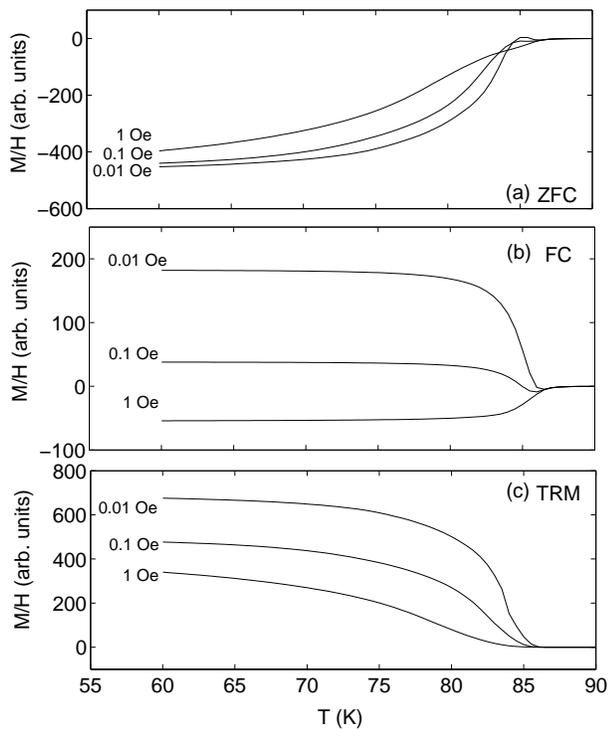}
\caption{Temperature dependence of the normalized (a) ZFC, (b) FC and (c) TRM magnetization for different applied magnetic fields; $H$=0.01, 0.1, and 1 Oe.}
\label{fig1}
\end{figure}

Figure~\ref{fig1} shows the temperature dependence of the ZFC, FC and TRM magnetization of Bi2212, recorded on re-heating using different magnetic fields. The different curves superpose as $M_{\rm FC}$-$M_{\rm ZFC}$=$M_{\rm TRM}$ for all the magnetic fields employed. At low temperatures, the shielding is complete, and in an applied magnetic field of $H=1$~Oe, Bi2212 behaves as an ordinary superconductor, with a large diamagnetic response up to $T_c$. In smaller fields, a large PME is observed, as evidenced by the FC results in Fig.~\ref{fig1}(b) and the ZFC data in Fig.~\ref{fig1}(a) at temperatures close to $T_c$. The low temperature PM state exhibits, as mentioned above, dynamical features typical to those of spin glasses. ZFC relaxation experiments are useful to investigate aging properties in spin glasses.\cite{relaxdc}
The relaxation of the low-frequency ac-susceptibility can also be recorded, and employed to investigate e.g. memory effects\cite{relaxac} in glassy materials. When cooling from above the spin glass transition temperature $T_g$, a halt at constant temperature $T_s$ below $T_g$ is made during $t_s$, allowing the system to relax to-wards its equilibrium state at $T_s$; this equilibrated state becomes frozen in on further cooling, and is retrieved on reheating. In the case of the glassy PM state of Bi-2212, the effect of the aging on the dc\cite{evieprl} relaxation is small and consequently also the relaxation of the ac\cite{evieejb} susceptibility is small, which makes difficult the above described ac-memory experiments.\cite{note}
``Dc-memory'' experiments, using a dc magnetic field (which in our measurement system yields a more stable signal than the ac-experiments) can instead be performed. Such experiments include single or multiple stops at different temperature below the critical temperature of Bi-2212.  
During these halts, the applied magnetic field value is kept constant (i.e. zero in the ZFC case and $H$ for the TRM). The validity of the method has been justified using conventional spin glasses,\cite{memoagmn,memoising} and it has been used to enlighten fundamental differences between different glassy systems.\cite{pero}

\begin{figure}[ht]
\includegraphics[width=0.45\textwidth]{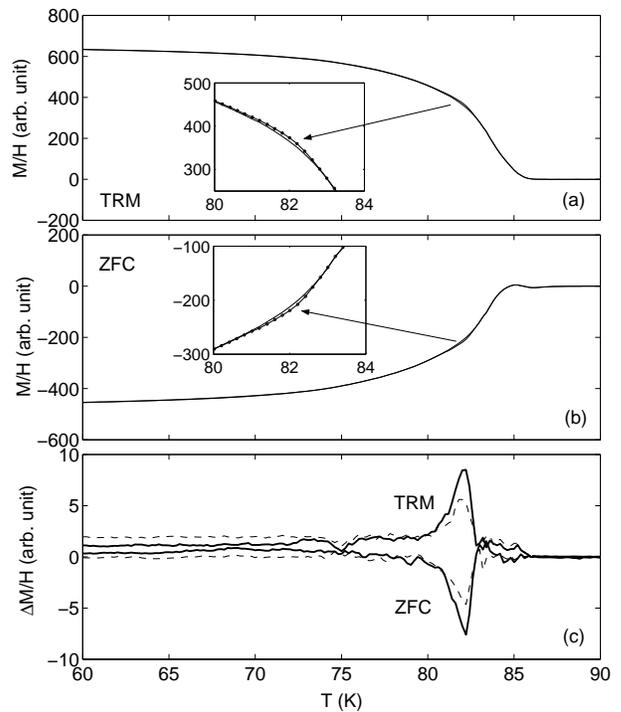}
\caption{Temperature dependence of the (a) TRM and (b) ZFC magnetization recorded on re-heating after direct cooling (simple line) and a temperature stop of $t_s$=10 000 s at $T_s$=82 K (markers); $H$=0.01 Oe. The inserts enlarge a smaller temperature region around the stop temperature. (c) Difference plots of the above curves ( continuous line ), the results for a 1 000 s  stop are added (dotted line).}
\label{fig2}
\end{figure}

Dc-memory experiments are performed using the temperature dependence of the ZFC and TRM magnetization. Figure~\ref{fig2}(a) and (b) shows the influence of a 10 000 s temperature stop at $T_s$ = 82 K on the ZFC (resp. TRM) curve. The ZFC (resp. TRM) curve corresponding to the stop experiment lies below (resp. above) a reference curve measured with the same cooling and heating rates (cf. corresponding experiments in ref. [7]). In both cases, as in ordinary spin glasses, the ``spin'' configuration established during the equilibration at $T_s$ during the stop has been frozen in on further cooling and is retrieved on re-heating. As seen in Figure~\ref{fig2}(c), which shows the difference plots of two pairs of magnetization curves, the 'dips' corresponding to the relaxation of the magnetization at $T_s$ are small in magnitude [$\sim 3 $\% of $M(T_s= 82 {\rm K})$], but well defined.

\begin{figure}[ht]
\includegraphics[width=0.45\textwidth]{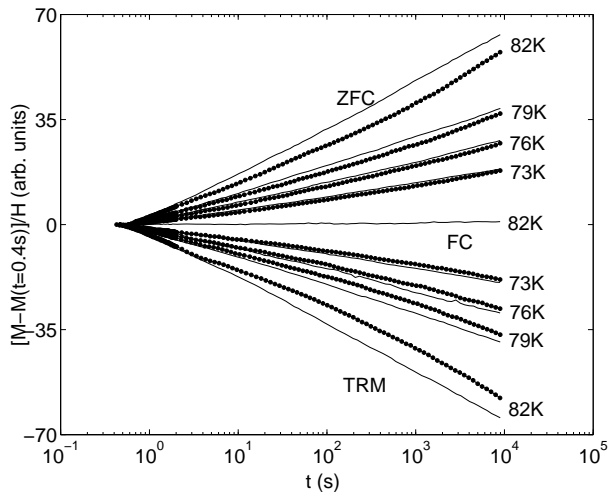}
\caption {ZFC and TRM relaxation curves recorded at different temperatures: $T_m$=73 K, 76 K, 79 K, and 82 K using $t_w$=0s (simple line) and $t_w$=10 000s (markers); $H$=0.01 Oe. A FC relaxation curve is added for $T_m$=82 K and $t_w$=0s.}
\label{fig3}
\end{figure}

Figure~\ref{fig3} shows the relaxation of the ZFC, FC, and TRM magnetization at different temperatures. As in conventional spin glasses, the FC relaxation is small,\cite{fcrelax,memoagmn} and for all temperatures the equality $M_{\rm ZFC}(t)-M_{\rm ZFC}(0.4 \; {\rm s})$ = - [$M_{\rm TRM}(t)-M_{\rm TRM}(0.4 \; {\rm s})$] is closely satisfied. The relaxation of the magnetization becomes smaller with decreasing temperature and the relaxation is larger for the $t_w=0$ s curves than for the $t_w=10\, 000$~s curves, revealing non-equilibrium dynamics - the system ages. Also, it is seen that the influence of the aging decreases with decreasing temperature. To relate these ZFC and TRM relaxation curves with the memory curves in  Figure~\ref{fig2}(a) we look specifically at the ZFC relaxation curves measured at $T_m =82$~K. It is seen that at an observation time of $\sim 30$~s (corresponding to the effective observation time of the magnetization 
measurement on heating), the relaxation curve measured directly after reaching $T_m$ lies significantly above the curve measured after a wait time of 10 000 s, just as the magnetization curve without a stop lies above the magnetization curve measured with a stop in Figure~\ref{fig2}(a). A closer examination even shows that the difference between the magnetization observed at 82 K in the dc-memory curves is closely equal to the difference between the ZFC relaxation curves at an observation time of 30 s for the $t_w=0$~s and the $t_w=10 \, 000$~s curves. These dc-memory results have their correspondence in ac-memory experiments. The level of the out-of-phase component of the ac-susceptibility is governed by the cooling rate employed in the experiments, yielding an ``effective age'' of the system. The susceptibility decays in magnitude during a stop at constant temperature,\cite{relaxac} as the system equilibrates and this lower level is retrieved on re-heating the sample.

As seen in Fig.~\ref{fig2}(c), the influence of a stop at a temperature $T_s$ = 82 K extends to neighboring temperatures, between 80 and 84 K. Outside this range the magnetization curve recorded after a stop is nearly identical to the reference one. A similar feature is observed in    ac-measurements, in which the susceptibility recovers its cooling-rate governed level on further cooling. It is common in this case to speak about rejuvenation, since the system appears younger, or reinitialized, at lower temperature. Using this analogy, one can also speak about rejuvenation in the dc-memory experiments.

\begin{figure}[ht]
\includegraphics[width=0.45\textwidth]{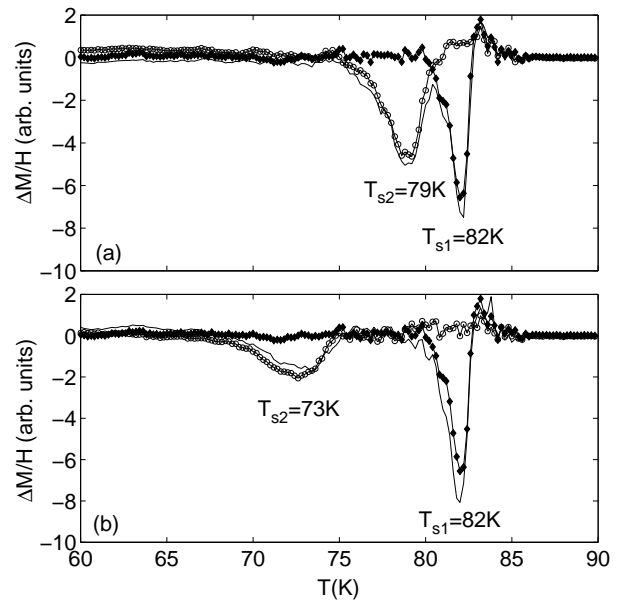}
\caption{(a) Difference plots corresponding to single stops of 10 000 s at $T_s$=82 K, $T_s$=79 K and to a double stop at both temperatures ($t_{s1}=t_{s2}$=10000s). (b) Idem with $T_{s1}$=82K and $T_{s2}$=73K.}
\label{fig4}
\end{figure}

\begin{figure}[t]
\includegraphics[width=0.45\textwidth]{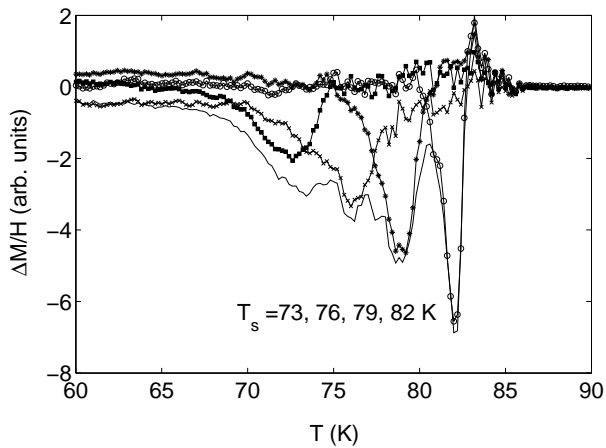}
\caption{Difference plots corresponding to single stops of 10 000 s at $T_s$=82 K, 79 K, 76 K and 73 K, and to a ``quadruple'' stop at the four temperatures ($t_{si}$=10000s).}
\label{fig5}
\end{figure}

The rejuvenation phenomena indicate that, outside an overlap region, the equilibration of the system at a temperature below the glass transition has no influence on the spin configuration at another temperature. Due to the chaotic nature\cite{bray,petra} of the glassy phase, the equilibrium state at one temperature is distinct from the one of a higher temperature, so that the relaxation at the lower temperature is not affected by the first equilibration at the higher one.\cite{note2} This is illustrated is Fig.~\ref{fig4}, which shows the difference plots of magnetization curves recorded after two stops at different temperatures. In Fig.~\ref{fig4}(a), the sample is first cooled to $T_{s_1}$ = 82 K, and after a 10 000 s stop, cooled to $T_{s_2}$ = 79 K where a second 10 000 s stop is performed. As in the single stop experiments, the cooling is then resumed to the lowest temperature and the magnetization (the ZFC magnetization in this case) is recorded on re-heating. 
The results of only  single stops at $T_s = 82$ K and $T_s = 79$ K are added for comparison. As mentioned above, the aging at $T_{s_1}$ is not influenced by the second equilibration at $T_{s_2}$, and vice-versa, the equilibration at $T_{s_2}$ is not affected by a first aging at $T_{s_1}$. These observations can be interpreted to reveal temperature chaos,\cite{bray} similar to that observed in conventional spin glasses.\cite{relaxac,memoagmn,memoising,pero,petra,vince} 
The two ``memory dips'' are sharp, and nearly distinct. If the second temperature stop is performed at a lower temperature, as in Fig.~\ref{fig4}(b), the two dips are completely separated; similar quite sharp and separated memory dips have also been observed in Cu(Mn) and Ag(Mn) glasses.\cite{relaxac,vince}
In the Ising case,\cite{memoising} the dips are much broader, and the overlap between neighboring temperatures seems larger than in the Heisenberg-like spin glasses.\cite{vince}
As seen in Fig.~\ref{fig5}, the dips are so sharp that multiple stops can be performed (4 in the present case), and all equilibrations can be retrieved on re-heating. It is interesting to note that the chiral glass model proposed by H. Kawamura \cite{kawamura1,kawamura2} can be used to describe both the glassy phase of the PME phase of our Bi-2212 sample and the Heisenberg spin glass.\cite{takayama}
A small amount of anisotropy may indeed exist in Heisenberg spin glasses, due to dipolar interaction and the scattering of the charge carriers by non magnetic impurities (Dzaloshinsky-Moriya interaction). In this case, the presence of frustration will induce a spin chirality,\cite{villain} and as for the PME sample, the frustration will be affected by the chirality of the non co-planar spin configuration.\cite{kawamura1}

The occurrence of a positive field cooled magnetization (PME) in superconductors is a phenomenon that is driven by different mechanisms in different systems.\cite{Gam,Ric} 
In Bi-2212 the PME is likely to arise from the existence of $\pi$-junctions and supercurrent loops with odd numbers of such junctions. In our current melt cast Bi-2212 system, the existence of spontaneous aging must necessarily relate to a frustrated collective state, which is created from a network of interacting 0- and  $\pi$-junction. The phenomenology of the aging behavior observed for the current Bi-2212 sample has some striking similarities with that of Heisenberg spin glasses, both of which systems may be pictured by a chiral glass model.\cite{kawamura1}

Financial support from the Swedish Research Council (VR) is acknowledged. A.G. acknowledges the ISP of Uppsala University for administrating and the Ministry of Science, Technology and Environment of Thailand for financing a fellowship for research in Sweden.

\end{document}